# Controllable light capsules employing modified Bessel-Gauss beams

Lei Gong,[1,*] Weiwei Liu,[1,*] Qian Zhao,[1] Yuxuan Ren,[2] Xingze Qiu,[1] Mincheng Zhong[1] & Yinmei Li[1]

[1]*Department Department of Optics and Optical Engineering, University of Science and Technology of China, Hefei, 230026, China*

[2]*National Center for Protein Sciences Shanghai, Institute of Biochemistry and Cell Biology, Shanghai Institutes for Biological Sciences, CAS, Shanghai, 201210, China*

*\*These authors contributed equally to this work.*
*Correspondence and requests for materials should be addressed to Y.M.L. (Email: liyinmei@ustc.edu.cn)*

**We report, in theory and experiment, on a novel class of controlled light capsules with nearly perfect darkness, directly employing intrinsic properties of modified Bessel-Gauss beams. These beams are able to naturally create three-dimensional bottle-shaped region during propagation as long as the parameters are properly chosen. Remarkably, the optical bottle can be controlled to demonstrate various geometries through tuning the beam parameters, thereby leading to an adjustable light capsule. We provide a detailed insight into the theoretical origin and characteristics of the light capsule derived from modified Bessel-Gauss beams. Moreover, a binary digital micromirror device (DMD) based scheme is first employed to shape the bottle beams by precise amplitude and phase manipulation. Further, we demonstrate their ability for optical trapping of core-shell magnetic microparticles, which play a particular role in biomedical research, with holographic optical tweezers. Therefore, our observations provide a new route for generating and controlling bottle beams and will widen the potentials for micromanipulation of absorbing particles, aerosols or even individual atoms.**

## Introduction

The rapidly growing optical tweezers techniques offer a precise and controllable access to manipulation of objects on micro- and nano-scale, resulting in many significant insights into events and processes on microscopic level occurring in biological, physical and chemical worlds[1]. Conventional optical tweezers confine objects in high-intensity region by counterbalancing scattering and gradient forces[2,3], so the captured objects are usually restricted to transparent particles with refractive index greater than that of the surrounding medium. However, the optical manipulation of absorbing particles or low-index particles still remain a challenge though many attempts have been made[4-6], holding back the increased interest of optical tweezers in relevant applications. Fortunately, a novel trapping configuration based on optical bottles or light capsules, composed of a huge darkness volume surrounded by higher intensity light barriers, can readily enable a stable confinement and guiding of light-absorbing particles[7-12]. Inside the light capsules, absorbing particles are repelled by high-intensity barrier and pushed towards regions of lower light intensity due to the thermal photophoretic force[7,13]. This will minimize the optical damage induced by light heating compared with the particles confined at high intensity region, which will be of particular interest to the *in-vivo* cell trapping experiments[14,15]. In addition, the optical bottle beams have also been exploited to capture cold atoms with decreased photon scattering and decoherence rates[16-20], low-index particles[21] and even chiral microparticles[22]. Furthermore, the concept of bottle beams is extended to areas of acoustic waves[23] and plasmonic surface beams[24-26]. All these expansions will open new avenues to applications of biomedical imaging, object manipulation and cloaking.

Over the past years, several optical schemes have been proposed to realize light capsules. Some make use of the coherent superposition of two vortex beams[11,20], Gaussian beams[18], Laguerre-Gaussian beams[27,28], or Bessel beams[29], and some take the advantage of moiré-techniques[10], speckle fields[30] or abruptly autofocusing beams[31]. They either require fine alignment of the optical system or lack the versatility. All these methods mentioned above are concentrated on scalar light fields, while the polarization capacity of light recently has also been adopted to create vectorial bottle beams with controllable light distributions[32,33]. This approach can adjust the bottle shape, but the main problem is that the generation of spatially variant vector optical fields is not a trivial task[34-36].

Here, we propose a novel optical scheme to generate controllable light capsules directly employing the intrinsic properties of modified Bessel-Gauss beams. Theoretically, we find that as long as the parameters of modified Bessel-Gauss beam are properly chosen, the light maxima are capable of self-bending along curved paths during propagation in free space, naturally creating a perfect darkness surrounded by regions of high light intensity, i.e. the light capsule. The light capsule is able to exhibit various geometrical shapes according to different choice of beam parameters. Remarkably, distinguished from the previous methods, not only the volume but also the darkness degree of the light capsule could be continuously regulated through tuning the beam parameters. For the experimental demonstration, a flexible scheme based on a digital micromirror device (DMD) with well-designed binary amplitude masks is proposed to shape the bottle beams by precise amplitude and phase manipulation. Besides, the fast switchable merit of the DMD enables the rapid switch among various bottle beams, which maybe benefit the dynamic applied scenarios. Finally, in order to demonstrate the specialties of the light capsules, we experimentally achieved a non-contact optical trapping and manipulation of core-shell magnetic microparticles using a liquid crystal spatial light modulator (LCSLM) based holographic optical tweezers[1,37]. Especially, the high controllability of the darkness degree and shape of the light capsules for the particle confinement were investigated, which can be exploited to capture and sort particles with different absorbing properties or refractive index. Thus, this technique may be beneficial for biomedical research due to the unique advantages of magnetic particles in applications of diagnosis, *in vivo* imaging, and drug delivery[38-40].

**Results**

**Principle and numerical simulation.** A detailed insight into the theoretical origin and characteristics of bottle beams employing modified Bessel-Guass beam is presented. The modified Bessel-Gauss mode is a special case of generalized Bessel-Gauss mode that can be derived from the superposition of decentered Hermite-Gaussian beams (see Methods). The modified modes ( $r_{d0} \neq 0, \varepsilon_0 = 0$ ) can naturally form optical bottles on condition that the beam parameters are properly selected. In free space (*A=D=1, B=z, D=0*), apart from the constant factor, the zero-order *modified* Bessel-Gauss beam has the form of

$$V(r,\theta,z) = \frac{\exp(ikz)}{\omega_0\sqrt{1+\xi^2}} \exp\left(\frac{\xi i - 1}{1+\xi^2}(R^2 + \sigma^2)\right) J_0\left(-\frac{\xi + i}{1+\xi^2} 2\sigma R\right) \quad (1)$$

with

$$R = r/\omega_0, \xi = z/L, \sigma = r_{d0}/\omega_0, L = k\omega_0^2/2. \quad (2)$$

In order to generate the optical bottle, there must be a huge darkness volume surrounded by high intensity light barriers. That means an intensity minimum should appear not only in the transverse pattern but also over propagation. Therefore, the transverse and on-axis intensity profiles in terms of the radial and longitudinal coordinates are analyzed to obtain the conditions.

In the waist plane, the intensity distribution of the modified beam is defined by

$$I(R) = A_0 \exp\left[-2(R^2 + \sigma^2)\right] I_0^2(2\sigma r), \tag{3}$$

where $I_0(x)$ indicates the zero-order *modified* Bessel function. Note that the transverse intensity profile is determined by the parameter $\sigma$. This geometric parameter regulates the intensity distribution. The intensity maximum stays in the center with geometric parameter smaller than 1, while the intensity peak split into two side lobes by increasing the geometric parameter. A supplementary animation (see Supplementary Movie 1) dynamically demonstrates how the intensity profile is regulated by the geometric parameter. These conclusions can be obtained mathematically through analysis of inflection point of Eq. (3). However, from the view of physics, this can be easily understood because such a beam is regarded as a superposition of Gaussian beams with mean wavevectors parallel to the longitudinal $z$ direction, whose centers are placed on a circumference of radius $\omega_0$ around the $z$ axis. By increasing $\sigma$, the maxima of the constituent beams recede from each other, and correspondingly the central intensity of the resultant field will change gradually from peak to valley. As a consequence, the radio of radius $r_{d0}$ and $\omega_0$ determines the transverse patterns. Fig. 1(a) demonstrates three typical cases of different values of $\sigma = 0.5, 1$ *and* $2$, which indicates the existence of a ring-structured pattern.

As for the on-axis intensity distribution with the form of

$$I(\xi) = \frac{1}{1+\xi^2} \exp\left[-2\sigma^2 \frac{1}{1+\xi^2}\right], \tag{4}$$

another animation (see Supplementary Movie 2) is presented to investigate the profiles in terms of the parameter as well. Taking the derivative of $I$ with respect to $\xi$, it can be observed that there must be a minimum of the intensity in the origin if $\sigma > \sqrt{2}/2$. On the other hand, if $\sigma < \sqrt{2}/2$, we have only a maximum point at the origin. Typical behaviors are verified by the results in Fig. 1(b), and thereby a nearly perfect null region can be clearly formed along the optical axis with large $\sigma$. As long as $\sigma > 1$, the beam will exhibit an intensity minimum in the origin but immediately increase the side lobe intensity (Fig. 1(c)), naturally creating an optical bottle. Further, as can be seen from the 3D model shown in Fig. 1(d), the optical bottle has a potential of saddle shape. Consequently, it is verified that the zero-order modified Bessel-Gauss beam is able to form three-dimensional light capsule enclosing supersized darkness when the initial parameter is properly chosen.

Generally, the control of the null field can provide optical capsules more possibilities for practical applications. However, in previous reports, only the size control can be realized usually through changing the NA of the focusing lens, so it is restricted to the capacity of optical elements and difficult to achieve the continuous adjustment. Notably, based on our method, not only the size but also the darkness degree of the light capsules can be continuously controlled by adjustment of a single parameter. As illustrated in Fig. 2(a), we define the radial size (*D*) and the axial size (*d_0*) of the null field in optical capsule as the distance between the two intensity maxima along radial and longitudinal direction, respectively. Fig. 2(b) presents the theoretical results about the dependence of the darkness region size on the parameter $\sigma$. Obviously, the radial and axial sizes of the null field increase in accordance with the parameter. As the parameter can be any positive real number, the darkness volume thereby can be manipulated continuously. In particular, the size of the optical capsule can be gigantic for large $\sigma$, implying that supersized null field can be created. Meanwhile, we find that not only the size but also the darkness degree, the degree of center intensity approaching zero, changes with the parameter. Especially for small $\sigma$, the intensity minimum in the origin descends fast with the increasing value, but for large $\sigma$, the intensity minimum gravitates towards zero and keeps unchanged basically, which is verified by the theoretical predictions in Fig. 2(c). Note that we could manipulate the null field through tuning the Gaussian beam waist $\omega_0$ that is in proportion to the radial size of the bottle. Therefore, it implies that our method can give rise to controllable light capsules enclosing perfect supersized darkness.

Now, let us consider the high order cases. In these cases, the transverse profile of the modified Bessel-Gauss beam defined by Eq. (5) is determined by the radial index *m* and azimuthal index *l*. As for $l=0$ but $m \neq 0$, the optical bottles can also be generated but under different conditions, for example, $\sigma > \sqrt{2}$ and $\sigma > \sqrt{3}$ in terms of $m=1$ and 2, respectively. These conditions can be obtained through analysis of the inflection point at $r=0$. Along with the increasing *m*, the beam will exhibit multiple rings (*m+1*) in the waist plane. Thus, the optical bottle generated will possess multilayer structures, and this will decrease the energy of the highest light barrier around the null field, as demonstrated in Fig. 3(a).

While in the case of $m=0$ but $l \neq 0$, the field has a phase vortex indicated by Eq. (8), resulting in a hollow-bottle beam, as illustrated in Fig. 3(b). Furthermore, the value of the index $l$ can change its shape down to the dark-hollow optical beams that is depicted in Fig. 3(c). When the two index are nonzero, the optical bottle beams will present more complex structures. Fig. 3(d) shows a typical one for the parameters of $m=1$ and $l=1$. Moreover, we note that the index $m$ also influences the size of the null field, providing another way to control the shape of the light capsules. So far, we have theoretically demonstrated that the *modified* Bessel-Gauss beams are able to naturally create three-dimensional bottle-shaped region during propagation as long as the parameters are properly chosen. Remarkably, the light capsules can exhibit various geometries by different choice of beam parameters. Further, the shape of the darkness can be continuously controlled by tuning the initial parameters, leading to adjustable light capsules.

**Bottle beam generation with a binary DMD.** To experimentally verify our findings, we demonstrate the generation of the bottle beams with different shapes. For the experimental scheme, a programmable digital micromirror device (DMD) is utilized for wavefront engineering. In particular, its capacity of simultaneous amplitude and phase modulation[41-44] is exploited for shaping the bottle beams. As the DMD is a binary SLM, binary amplitude masks are first designed based on a super-pixel method to accurately encode both the amplitude and phase information[43] (see Methods). Fig. 4(c) illustrates a typical binary DMD pattern that encodes the corresponding amplitude and phase information, depicted in Figs. 4(a) and 4(b), of a modified Bessel-Gauss beam (*m=1, l=1*). Practically, encoding is achieved by applying a sequence of ON and OFF states to the DMD that lie on the optical path of each user-defined region of interest. Our setup is presented in Fig. 4(d). The employed DMD (0.7 inch, XGA, Texas Instruments) is located at the front focal plane (image plane) of the 4-*f* configuration composed of two lenses (L3=250mm, L4=100mm), and a pinhole placed at the Fourier plane selects only the beam of first diffraction order and rejects the components with high spatial frequencies. Once the surface of DMD projected with predesigned holograms is sufficiently illuminated by a collimated laser beam ($\lambda=532nm$), the desired field distribution will appear at the back focal plane. Finally, a movable CCD camera allows us to record the intensity profiles at different positions along propagation direction, upon which the 3D light capsules can be subsequently reconstructed.

First, we demonstrate the generation of various Bessel-Gauss modes with our proposed scheme. The corresponding binary patterns are designed with super-pixel method in advance and projected onto the DMD. Subsequently, the intensity images of various generated modes can be observed at the target plane, illustrated in Figs. 4(e)-4(h). Beneath these pictures, we also present the corresponding one-dimensional transverse intensity profiles reconstructed at *y=0*. An excellent match can be seen compared with theoretically calculated profiles which are shown in Figs. 4(i)-4(l). Notably, the dynamic generation of such modes can be realized with our DMD-based scheme, which has the highest switching frequency of 22 kHz in principle[45]. This may benefit the dynamic applied scenarios.

To obtain the three-dimensional light capsules, a series of cross-sections along longitudinal axis were collected near the target plane. Figs. 5(a)-5(c) illustrate three individual transversal field intensities with equal separation distances of zero-order modified Bessel-Gauss mode, while Figs. 5(e)-5(g) demonstrate those of the high-order mode with $m=1, l=1$. Accordingly, the *x-z* distributions of reconstructed light capsules are presented in Fig. 5(d) and Fig. 5(h), respectively. The nearly perfect darkness surrounded by regions of high light intensity are clearly distinguished in the measured distributions. We can see that not only bottle beams but also hollow-bottle beams are demonstrated in our experiments. Remarkably, this can be easily achieved through tuning the beam parameters. Therefore, the advantages of such bottle beams and the DMD-based scheme have been combined to enhance the controllability and flexibility for generation of light capsules.

**Optical trapping of core-shell magnetic microparticles.** To demonstrate the specialties of the light capsules, we realize a non-contact optical trapping and manipulation of core-shell magnetic microparticles using a LCSLM based holographic optical tweezers (see Methods). Magnetic particles play a particular role in biomedical applications, such as diagnosis, *in vivo* imaging and drug delivery[38-40]. They are normally moved under a magnetic field, but cannot be trapped and manipulated to a targeted position. Fortunately, optical tweezers that enable manipulation of small particles provide a promising tool, but it is difficult to trap magnetic particles with a Gaussian beam because of their strong absorbing property. Here we take advantage of the intrinsic properties of the light capsules and its thermal photophoretic force on absorbing particles in liquid to achieve the optical manipulation of magnetic microparticles.

Figure 6 presents the results of optical trapping and manipulation experiments together with two supplementary videos (see Supplementary Movie 3 and 4) that show the dynamic process. The magnetic particles (SiO2@Fe3O4 core-shell microspheres, 4.0–5.0 μm in diameter, BaseLine Company) suspended in a chamber filled with water are used in our experiments. They can be trapped inside light capsules created by *modified* Bessel-Gauss beams. As a demonstration, we employ the zero-order mode for trapping experiments. As depicted in Fig. 1(d), the different length scales in the transverse and axial direction (about 8 μm and 25 μm in experiment) underline the asymmetric bottle shape, leading to lower confinement in the axial direction compared with that in transverse plane. So when we move the trapped particle by moving the chamber, it will be not very stable inside the light capsule because of perturbation, and that's the reason why defocus blur is observed in the process. Meanwhile, the particle is also pulled off the center of the optical bottle in the transverse plane and maybe partially go beyond its soft boundary caused by viscous drag force of the fluid. Nevertheless, it cannot escape the optical bottle through its bottlenecks and still be confined inside the capsule. Hence, we can conclude that the light capsules are able to confine and manipulate core-shell magnetic particles in three dimension. Actually, we recently have demonstrated optical trapping of such particles with cylindrical vector beams and particularly exploited their heating effects to kill targeted cancer cells[46]. However, there are lots of applied researches, especially the *in vivo* cases, where heating effects are really undesired, while light capsules can make a difference because the particles are confined at the low intensity region. Thus the noninvasive trapping technique will greatly benefit magnetic microparticles' application in biomedical research.

Further, we also investigated the high controllability of the darkness degree of the light capsules in the manipulation experiments. As demonstrated in Fig. 2, we can continuously adjust the light capsule especially the darkness degree that decides the potential depth of the optical trap. Note that potential depth can also be changed via laser power control, while here it is realized by controlling the intensity gradient inside the light capsules which essentially determines the trapping ability. In the experiments, bottle beams with different darkness degrees were employed to trap the same kind of light-absorbing particles. The first two rows of Fig. 7 demonstrate the confinement of particles using light capsules with $\sigma = 1.5$ and $\sigma = 2.5$ ($\sigma$ determines the darkness degree of the zero-order bottle beam), respectively. We find that the former light capsule had very weak force on the magnetic particle. When we shifted the particle a little away from the center, it moved slowly back towards the center (Figs. 7(a)-7(c)). Whereas the particle easily escaped from the trap once a little external force was exerted (Fig. 7(d)). While the latter could stably confine and manipulate the particles in three dimension (Figs. 7(e)-7(h)). This is because the potential of latter bottle is deep enough to confine such an absorbing particle. Similarly, the last row of Fig. 7 presents the images of a trapped particle with the first order beam (*m=0, l=1*, $\sigma = 2.5$) which possesses a phase vortex. Such bottle beam has a perfect darkness in the center, resulting in an enhanced confinement of the absorbing particles. This is evidenced by the smaller motion range indicated by the dashed circles compared with that in Figs. 7(e)-7(h) under the roughly same moving speed of fluid. Our observations demonstrate that light capsules with different darkness degrees have different confinement abilities in terms of the same light-absorbing particles. On the contrary, a certain bottle beam has different confinement abilities for different absorbing particles. That means critical trapping of particles with various absorbing properties or refractive index requires light capsules with controllable darkness inside. Thus, this can be exploited to trap and sort these kinds of particles. Notably, our high controllability of the darkness degree and shape of the light capsules may play a key role in these applications.

**Discussions**

In the theoretical section, we focus on the discussion of light capsules created by modified Bessel-Gauss beams. Here we will discuss the more generalized modes, figuring out the influence of parameter $\varepsilon_0$ upon the formation of bottle beams. Taking the zero-order generalized Bessel-Gauss beam as an example, for the given choice of the parameters ($\sigma = 2.5$ and $\varepsilon_0 = 10^o$), its propagation behaves differently compared with that of the modified beams. In Fig. 8(a), we show the transverse intensity profiles of the generalized beam at different propagation distances. As the distance increases, even though a ring-shape structure appears in the waist plane, the intensity will not increase quickly up to a much high value any more along the optical axis, which is distinguished from the behaviors of the modified beam presented in Fig. 1(c). In this regard, the generalized modes cannot be used to generate the optical bottles. However, from a mathematical point of view, the transition from the generalized modes to modified modes occurs when $\varepsilon_0 \to 0$. Therefore, it resembles the behaviors of the modified ones when $\varepsilon_0$ approximates to zero. Fig.

8(b) gives an example distribution of the generalized with $\varepsilon_0 = 0.01^o$, which is much similar with that of modified one illustrated in Fig. 1(d). The observation agrees with our theoretical predictions. Hence, we conclude that the generalized Bessel-Gauss beams can also create optical bottles in free space as long as the previous conditions are satisfied and simultaneously the parameter $\varepsilon_0$ approximates to zero.

In fact, such generalized modes can be regarded as quasi-modified ones. Next, we will investigate the influence of parameter $\varepsilon_0$ on the generated optical bottles. Fig. 8(c) shows the field distribution of the 3D light capsule with $\varepsilon_0 = 0.5^o$ while other parameters are the same with Fig. 8(b). Interestingly, we see that the intensity distribution of the light capsule created by generalized mode is not asymmetric in the propagation direction. And on the increasing value of $\varepsilon_0$, the asymmetric behavior becomes more and more evident. Further, when the parameter is assigned with a negative value, the asymmetric behavior will change over, which can be verified by the optical patterns in Figs. 8(c) and 8(d). This phenomenon can also be observed for light capsules formed by high-order generalized Bessel-Gauss beams. Note, that such asymmetry of the intensity distribution of the light capsules can result in asymmetric depth of potential well when used in optical trapping, which may expand the potentials of the light capsules, especially benefiting from the continuously adjustment of the asymmetric distribution.

In conclusion, we introduce and demonstrate a new class of light capsules enclosing perfect supersized darkness in free space. Theoretically, we show that light capsules with various geometrical shapes can be formed with *modified* Bessel-Gauss beams so long as the parameters are properly chosen. Notably, compared with the previous approaches, the geometrical shape and darkness degree of the light capsules can be tuned continuously by adjustment of the initial beam parameters. As a proof of concept, the generation of bottle beams with different shapes was demonstrated employing binary amplitude holograms projected onto DMD. Our technique brings about unique advantages in terms of flexibility and enhanced tunability. The experimental observations agree well with the theoretical analysis. Furthermore, to demonstrate the specialties of light capsules, we realize a three-dimensional optical trapping and manipulation of core-shell magnetic microparticles in liquid, which play an important role in biomedical research. And the high controllability of the darkness degree and shape of the light capsules were experimentally investigated to provide the possibility for sorting particles with different absorbing properties or refractive index. Finally, we discuss the special cases of light capsules derived from the generalized Bessel-Gauss beams. Note that, as generalized high-order Bessel-Gauss beams are paraxial solutions to the scalar Helmholtz equation, they can be implemented for acoustic waves or plasmonic surface waves as well. Hence, our observations provide new routes for generating and controlling bottle beams and the advantages of optical manipulation with controllable light capsules will greatly benefit applications in various fields.

**Methods**

**Theoretical formulation of generalized Bessel-Gauss beams.** The generalized Bessel-Gauss beam solution is derived from the superposition of decentered Hermite-Gaussian beams that are conventional Hermite-Gaussian beams with a displacement in the beam center by a vector $\mathbf{r_{d0}}$ in the *x-y* plane and a tilt of the mean beam direction with respect to the *z* direction by an angle $\varepsilon_0$ [47-49]. If the vectors $\mathbf{r_{d0}}$ and $\boldsymbol{\varepsilon_0}$ of the component beams are collinear with $r_{d0} = const.$ and $\varepsilon_0 = const.$, that means the major centers of component beams are placed on a circle with radius $r_{d0}$ in the *z=0* plane and their mean beam directions point to the apex of a single cone with a semi-aperture angle $\varepsilon_0$, the generalizing Bessel-Gauss beams are formed with the following form[49]:

$$V_{ml}(r,\theta,z) = \frac{2\pi}{\omega}\exp(ikz)\exp(-i\phi\cdot(m+1))\exp(i\varphi)\exp\left[\frac{ik}{2q}(r^2+r_d^2)\right]\mathscr{S}_{ml}, \quad (5)$$

where $q$ is the Gaussian beam parameter and $\omega$ is the Gaussian waist. They are related by $\omega = \sqrt{2/k/\mathrm{Im}(1/q)}$. The parameter $\varphi$ denotes a constant phase due to the transformation, while the other constant phase in terms of $\phi$ is the Gaussian Gouy phase. And the term $\mathscr{S}_{ml}$ is given by

$$\mathscr{S}_{ml} = \frac{1}{2\pi}\int_0^{2\pi} d\gamma H_m\big(a\cos(\theta-\gamma)-b\big)\cdot\exp\big(i\alpha\cos(\theta-\gamma)\cdot\exp(il\gamma)\big), \quad (6)$$

where

$$\alpha = kr(\varepsilon-(r_d/q)); a = \sqrt{2}r/\omega; b = \sqrt{2}r_d/\omega. \quad (7)$$

The parameters $r_d$ and $\varepsilon$ at any propagation distance can be related to the initial ones via ABCD propagation matrix: $r_d = Ar_{d0} + B\varepsilon_0$ and $\varepsilon = CAr_{d0} + D\varepsilon_0$.

To generalize the analysis, an azimuthal phase-variation is introduced for the integration,

which is characterized by the index *l*. The integral $\mathscr{I}_{ml}$ can be solved analytically, and for example the first three expressions are given by[49]

$$\mathscr{I}_{0l} = e^{il\theta} \cdot i^l \cdot J_l(\alpha), \tag{8}$$

$$\mathscr{I}_{1l} = \frac{e^{il\theta}}{2} \cdot i^l \cdot \left[ 2ia\left((-1)^l J_{1-l}(\alpha) + J_{1+l}(\alpha)\right) - 4bJ_l(\alpha) \right], \tag{9}$$

$$\mathscr{I}_{2l} = \frac{e^{il\theta}}{2} \cdot i^l \cdot \left[ -2a^2\left((-1)^l J_{2-l}(\alpha) + J_{2+l}(\alpha)\right) - 8iab\left((-1)^l J_{1-l}(\alpha) + J_{1+l}(\alpha)\right) \right. \\ \left. + \left(4a^2 + 8b^2 - 4\right)J_l(\alpha) \right]. \tag{10}$$

Obviously, the radial index *m* and azimuthal index *l* decide the field distribution, and further it depends on the scalar parameters $\omega_0$, $r_{d0}$, and $\varepsilon_0$. So the field patterns could be easily modulated through tuning these parameters. To simplify the discussion, we first classify such beams according to the parameters $r_{d0}$ and $\varepsilon_0$ and the field distributions at the waist plane[47,49]. For $r_{d0} = 0, \varepsilon_0 \neq 0$, the argument is $\alpha = k\varepsilon_0 r$ at the *z=0* plane, thus it reduces to the more familiar, ordinary Bessel–Gauss beam. While for $r_{d0} \neq 0, \varepsilon_0 = 0$, the argument $\alpha = -i2r_{d0}r/\omega_0^2$ is purely imaginary. Considering the equation $J_n(-i\beta r) = (-i)^n I_n(\beta r)$, the transverse pattern of the beam can be described by the modified Bessel function in the waist plane, thereby it is termed as *modified* Bessel-Gauss beam. Apparently, when both are nonzero, Eq. (5) represents the more generalized situations.

**Full field shaping with binary amplitude hologram.** We take the advantage of the binary amplitude masks[50] projected onto DMD to shape the bottle beams. It requires manipulation of the spatial full field information (i.e. amplitude and phase) of a light beam[51]. Thus, the binary hologram must be designed in advance to encode the complex signal. Usually, the Lee hologram is adopted to perform this task[52]. Here we employ the super-pixel based method where the neighboring pixels within hologram are grouped into various super-pixels to define a complex field in the target plane[43,44]. The Lee hologram only takes the advantages of one dimension to encode the amplitude and phase in fringes, while this technique exploits both dimensions of the pixel array using super-pixels, thereby providing more precise encoding.

The key of the super-pixel method is to realize the phase prefactors of the target plane responses of component pixels within each super-pixel are distributed uniformly between 0 and $2\pi$ by designing a configuration with DMD and a filter. As depicted in Fig. 4(d), the two lenses in the 4*f* configuration is placed a little off-axis with each other, and an aperture filter is adopted to obstruct the high spatial frequencies in the Fourier plane, so that the target plane response of a super-pixel is the sum of the individual pixel responses. Then one can construct a great many target fields by turning on different combinations of pixels within a super-pixel. Hence, the design of the binary masks is to determine the combination of on and off pixels within every super-pixel in terms of the desired field distribution. A lookup table can be employed to calculate the holograms more efficiently[43]. Finally, the bottle beams can be shaped with these binary amplitude masks loaded onto the DMD.

**Holographic optical tweezers setup.** To implement the particle trapping experiments, we constructed a holographic optical tweezers based on an inverted microscope (IX70, Olympus), which enables complex-amplitude modulation employing an off axis hologram[53]. The hologram displayed on an LCSLM (Holoeye PLUTO) is fully illuminated by an expanded Nd:YAG laser (HPG-5000, Elforlight, $\lambda = 532 nm$) for shaping the bottle beams. The first-order diffraction component is selected via a 4*f* telescope with a pinhole in the Fourier plane and then coupled into an oil-immersion objective (100X, numerical aperture 1.30, Olympus) through a lens (*f*=250*mm*). Near the focal plane, the light capsule is formed and subsequently used for trapping and manipulation of the magnetic particles with the help of a mechanical stage (M545, Physik Instrumente). The trapped particles can be observed clearly because the center of the light capsule is located near the imaging plane and their images were recorded by a CMOS camera (1.4 M100, DALSA). The sizes of the bottle beam could be estimated from the images reflected by the cover slip through axial adjustment of the microscope objective.

**Acknowledgements**

We thank Dr. Haowei Wang from University of Science and Technology of China for useful discussions about the optical trapping experiments. This work was supported by the National Natural Science Foundation of China (Grant Nos. 11374292, 61535011 and 11302220).

**Authors Contributions**

L. G. and Y. M. L. conceived the idea. L. G. and X. Z. Q. carried out the simulation. L. G. and Q. Z. designed and performed the free-space experiments. W.W.L. constructed the holographic optical tweezers and performed the trapping experiments. L. G. and Y. X. R. contributed to data analysis. L.G. prepared the manuscript, and M. C. Z contributed to figure preparing. Y. M. L. provided overall supervision. All authors were involved in revising the manuscript.

**Additional Information**

Additional supporting videos may be found in the online version of this article at the publisher's website.

**Competing financial interests:** The authors declare no competing financial interests.

**Figure Legends**

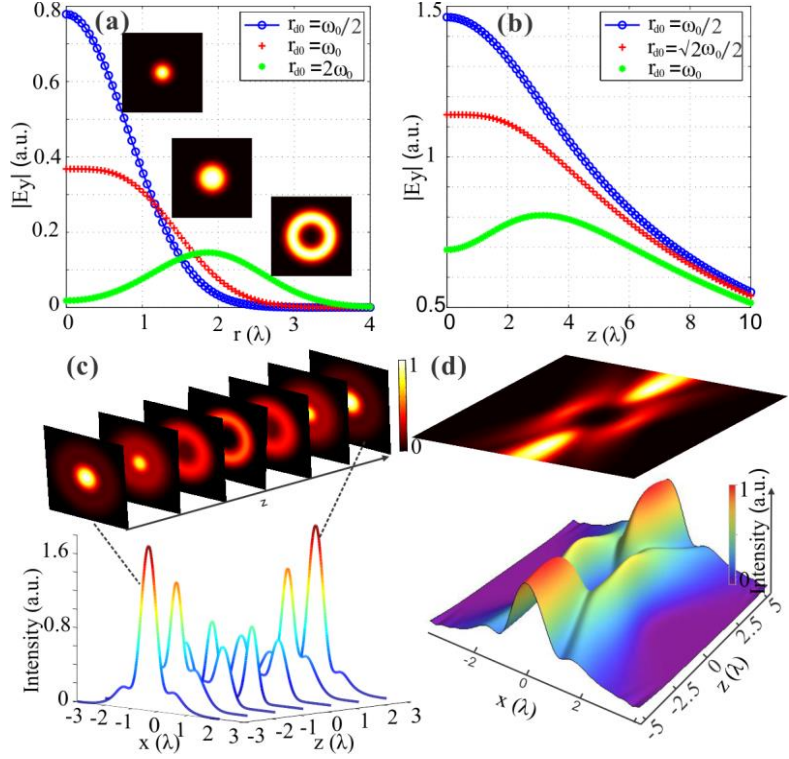

**Figure 1 | Formation condition of optical bottle employing zero-order *modified* Bessel-Gauss beam and the intensity distributions of light capsule.** **(a)** Transverse intensity profiles of zero-order *modified* Bessel-Gauss beams with different parameters of $\sigma = 0.5$ (blue line), 1(red line), and 2 (green line) and the corresponding 1D intensity curves. **(b)** Longitudinal intensity curves of the zero-order mode for three typical $\sigma$. **(c)** Seven cross-sections with equal separation distances along optical axis of the bottle beam ($\omega_0 = \lambda$ and $\sigma = 2$) together with the correspongding 1D transverse intensity profiles. **(d)** The *x-z* intensity profile of the light capsule formed by zero-order *modified* Bessel-Gauss beam and its 3D model (bottom).

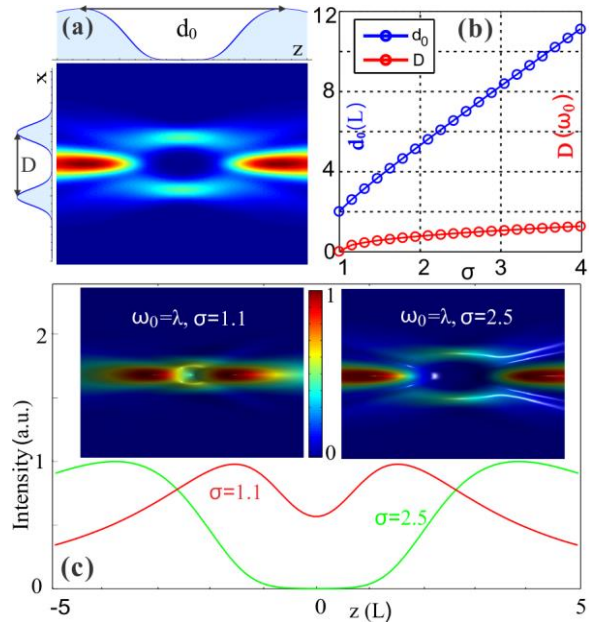

**Figure 2 | Control of the sizes and the darkness degree of the light capsule.** (a) Sketch definition of the radial and axial sizes of the light capsule. (b) The relationship between radial (red line) or longitudinal (blue line) size of the null field in optical capsule and the increasing parameter $\sigma$. (c) The control of the darkness degree inside the optical bottle. This can be verified by comparing the central intensity values of the on-axis intensity profiles for different cases ($\sigma=1.1$ and $\sigma=2.5$). The inset shows the intensity map of the corresponding two cases, in which the darkness degree inside the light capsules can be clearly observed.

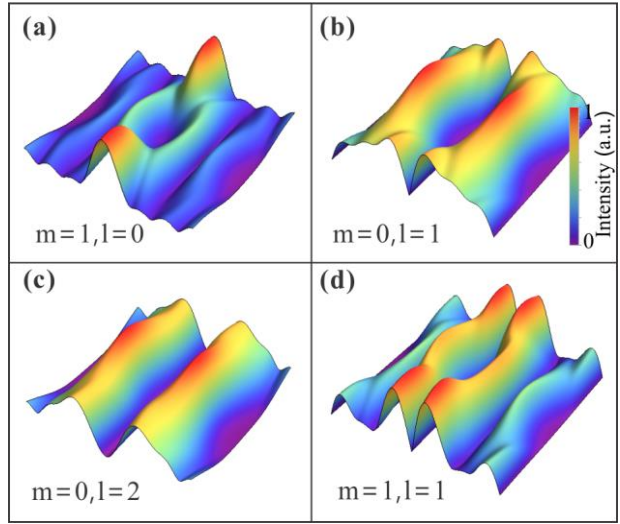

**Figure 3 | Light capsules with various geometrical shapes employing high-order *modified* Bessel-Gauss beams.** (a) 3D intensity distribution of the light capsule genterd with *modified* mode of *m=1* and *l=0*. (b)-(c) 3D optical bottles employing *modified* modes with the same *m=0* but with different *l* (*l=1 and 2*, respectively). (d) 3D distribution of the light capsule formed with mode of *m=1* and *l=1*. The beam parameter for all the calculated bottle beams are $\sigma=2.5$, $\omega_0=\lambda$ and $\lambda=532nm$.

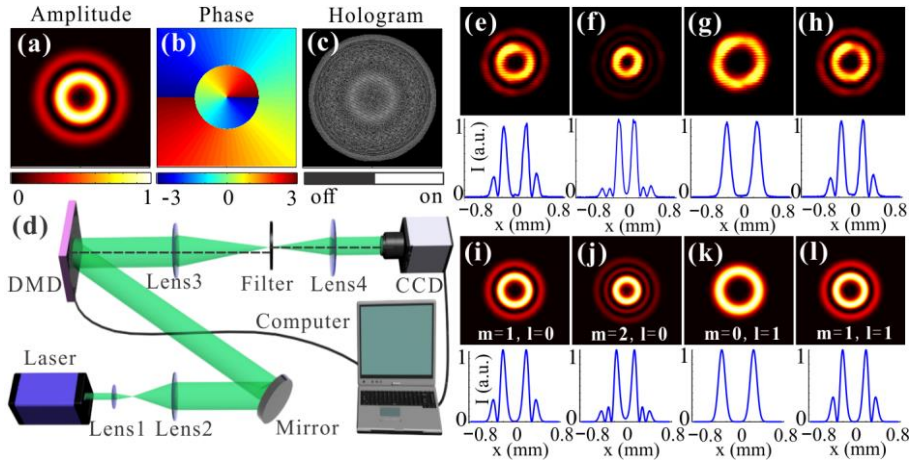

**Figure 4 | Experimental implements and measurement results for various *modified* Bessel-Gauss modes.** (a) Normalized amplitude and (b) phase of a *modified* Bessel-Gauss mode (*m=1, l=1*) and the corresponding binary DMD pattern (c) to generate such mode encoding the

desired field with super-pixel method. **(d)** Experimental setup. **(e)-(h)** Images of the generated Bessel-Gauss beams with various orders [(*m=1, l=0*), (*m=2, l=0*), (*m=0, l=1*), (*m=1, l=1*)] and the 1D intensity profiles reconstructed at *y=0*. **(i)-(l)** The calculated intensity distributions corresponding to those in Figures (e)-(h). The other parameters are the same for all the generated beams, which are $\omega_0 = 250\mu m$ and $\sigma = 2.5$.

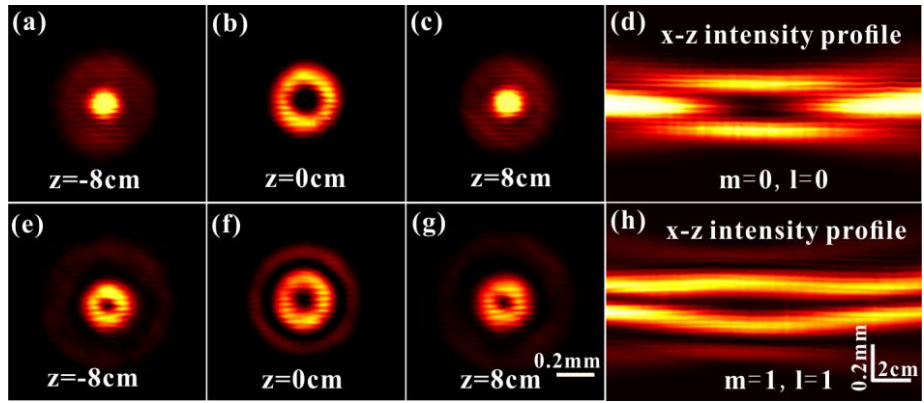

**Figure 5 | Experimental results of the light capsules. (a)-(c)** Three typical intensity cross-sections of equal interval near the imaging plane and **(d)** corresponding profile of the generated light capsule employing zero-order *modified* Bessel-Gauss beam. **(e)-(g)** The measured transverse intensity profiles of a high-order *modified* mode (*m=1,l=1*) along optical axis. **(h)** The *x-z* profile of the generated bottle-hollow beam. $\sigma = 2.5$ and $\omega_0 = 200\mu m$ in experiments.

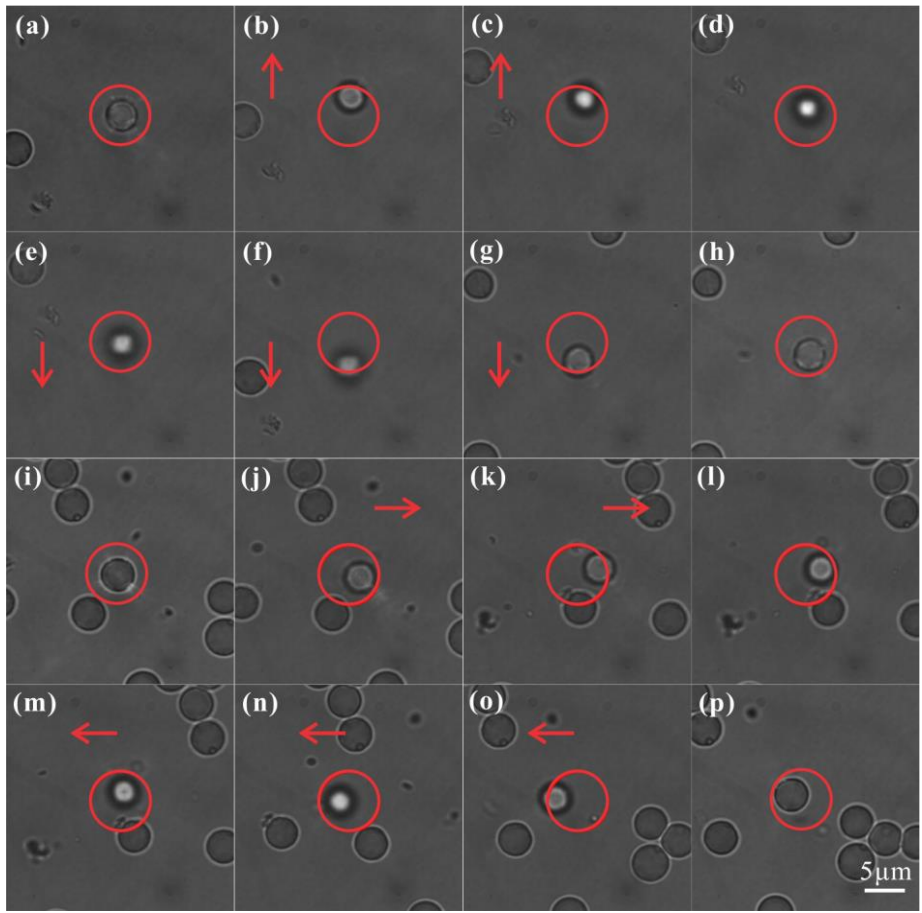

**Figure 6 | Optical trapping of core-shell magnetic microparticles with light capsules. (a)** The magnetic particle is trapped in the center of the light capsule first. When we move the chamber along the vertical direction, although the particles accordlingly deviate from the center upwards **(b)-(c)** or downwards **(e)-(g)**, they cannot escape the optical bottle eventually. Once moving stops, the particles will stay for a little time near the upper **(d)** or lower **(h)** boundary of the light capsule and then move towards the center. **(i)-(p)** The similar pocess of manipulating a magnetic particle but along horizontal direction. The circles indicate the range of the light capsules, and the arrows respresent the directions of the flowing fluid.

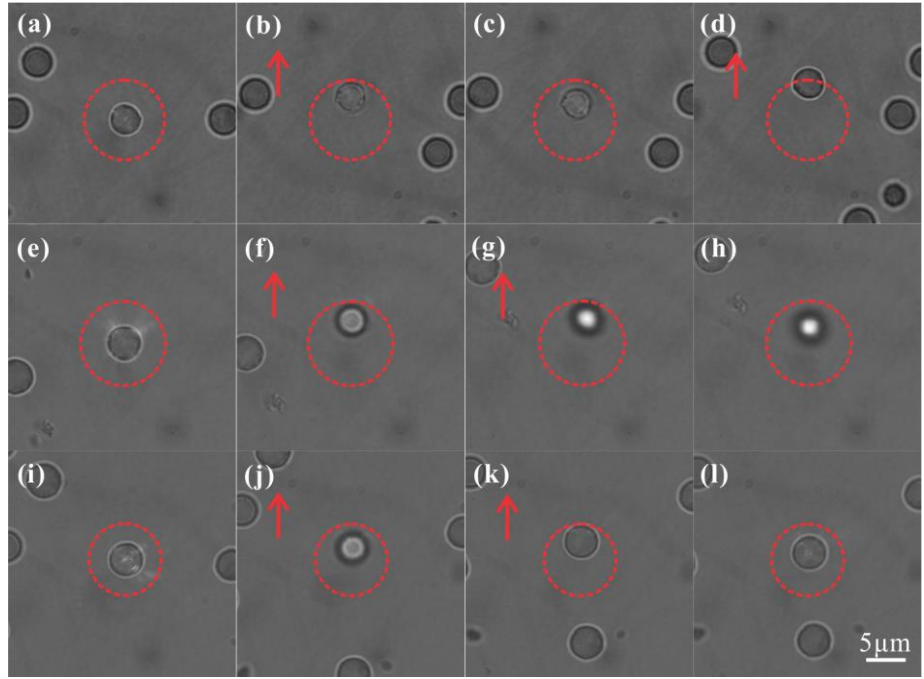

**Figure 7 | Optical confinement of absorbing particles using light capsules with different darkness inside. (a)**-**(c)** Trapping demonstration using light capsule with $\sigma=1.5$. Such bottle beam has very weak force on the particle. When we shifted the particle a little away from the center, it itsself moved slowly back to the center. **(d)** But with a little external force exerted by the flowing fluid, it easily escaped from the optical bottle. **(e)-(h)** The demonstration of stably trapping and manipulating a magnetic particle using bottle beam with $\sigma=2.5$. The particle could be confined within the capsule even with larger external force. **(i)-(l)** Enhanced confinement and manipulation of a magnetic particle employing the first-order bottle beam ($m=0$, $l=1$, $\sigma=2.5$). Compared with the previous trapping, the trapped particle has smaller motion range indicated by the dashed circles under a certain moving speed of the fluid. The arrows respresent the directions of the flowing fluid.

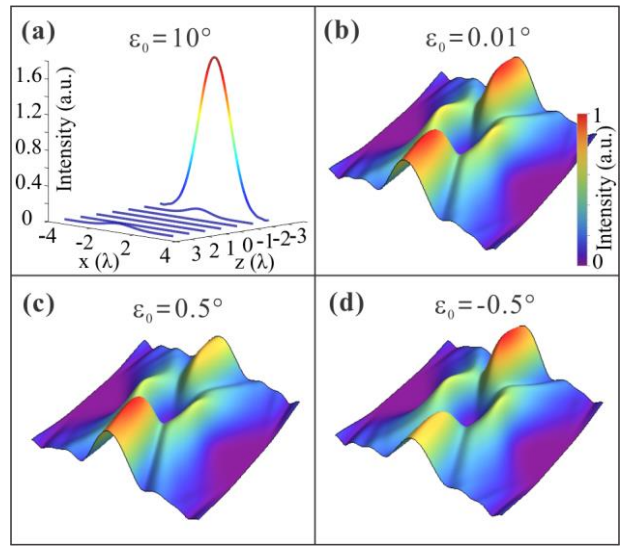

**Figure 8 | Intensity distributions of the generalized Bessel-Gauss beams under four sets of parameters that determine the conditions under which the light capsules can be formed.** **(a)** Transverse intensity profiles of the zero-order generalized mode with $\varepsilon_0 = 10^0$ at different cross-sections. **(b)** 3D intensity profile of the generalized mode with $\varepsilon_0 = 0.01^0$. **(c)-(d)** The asymmetric distributions of generalized beams with $\varepsilon_0 = 0.5^0$ and $\varepsilon_0 = -0.5^0$, respectively. In simulations, the beam parameters are set to be $\sigma = 2.5$ and $\omega_0 = \lambda$.